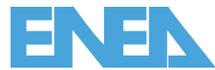

# THE FINE STRUCTURE CONSTANT AND NUMERICAL ALCHEMY

G. DATTOLI

ENEA -Tecnologie Fisiche e Nuovi Materiali, Centro Ricerche Frascati

# THE FINE STRUCTURE CONSTANT AND NUMERICAL ALCHEMY

*Abstract*

*We comment on past and more recent efforts to derive a formula yielding the fine structure constant in terms of integers and transcendent numbers. We analyse these "exoteric" attitudes and describe the myths regarding α, which seems to have very ancient roots, tracing back to Cabbala and to medieval alchemic conceptions. We discuss the obsession for this constant developed by Pauli and the cultural "environment" in which such an "obsession" grew. We also derive a simple formula for α in terms of two numbers π and 137 only. The formula we propose reproduces the experimental values up to the last significant digit, it has not any physical motivation and is the result of an alchemic combination of numbers. We make a comparison with other existing formulae, discuss the relevant limits of validity by comparison with the experimental values and discuss a criterion to recover a physical meaning, if existing, from their mathematical properties.*

**Riassunto**

In questo articolo si discutono i tentativi svolti in passato ed in epoche più recenti per derivare una formula che permetta di scrivere la costante di struttura fine in termini di numeri interi e trascendenti. Si analizzano tali tentativi che hanno radici antiche, rintracciabili nella Cabbala e nelle concezioni alchemiche medievali. Discuteremo l'*ossessione* per tale costante sviluppata da un fisico del calibro di Pauli e descriveremo l'ambiente culturale in cui essa maturò. Deriveremo anche una formula per α scrivibile solo in termini di 137 e π. La formula da noi proposta, priva di alcun fondamento fisico e ispirata a pure considerazioni numerologiche, riproduce i dati sperimentali fino all'ultima cifra decimale significativa. Si discute infine la possibilità di riconciliare tale risultato con le tecniche diagrammatiche di Feynman e con le più recenti stime analitico-numeriche della costante di struttura fine.

# INDICE

# THE FINE STRUCTURE CONSTANT AND NUMERICAL ALCHEMY

## 1. INTRODUCTION

There is a dream, which, albeit more often not confessed, occupies the most secret aspirations of theoreticians and is that of reducing the various "constants" of Physics to simple formulae involving integers (possibly primes) and transcendent numbers (essentially $e$ and $\pi$).

Within this context, the physical quantity which has created more interest is perhaps the fine structure constant $\alpha = (e^2/\hbar c)$ [1], originally introduced in Physics by Sommerfeld to include the relativistic corrections in the Bohr theory of the hydrogen atom [2]. Successively it became a key parameter in the quantum electrodynamics and it controls hyperfine splitting of the hydrogen atom spectral lines (see Fig. 1).

This quantity became, as we will see in the following, a real obsession for a great physicist like Pauli [1] and, about 25 years ago, Feynman used these inspiring words to describe a too often felt feeling of frustration, by who is trying to enter more deeply in the intimate nature of $\alpha$ [2].

"…**It has been a mystery ever since it was discovered more than fifty years ago, and all good theoretical physicists put this number up on their wall and worry about it. Immediately you would like to know where this number for a coupling comes from: is it related to $\pi$ or perhaps to the base of natural logarithms? Nobody knows. It's one of the greatest damn mysteries of physics: a magic number that comes to us with no understanding by man. You might say the 'hand of God' wrote that number, and 'we don't know how He pushed his pencil.' We know what kind of a dance to do experimentally to measure this number very accurately, but we don't know what kind of**

---

[1]  We use rationalised cgs units, therefore $4\pi\varepsilon_0 = 1$

[2]  Sommerfeld introduced the fine structure constant by noting that the reduced velocity (β=v/c) of the electron in the first Bohr orbit is just β=α and that the energy level formula can be written as $E_n = -\frac{1}{2}\frac{\alpha^2}{n^2}m_e c^2$.

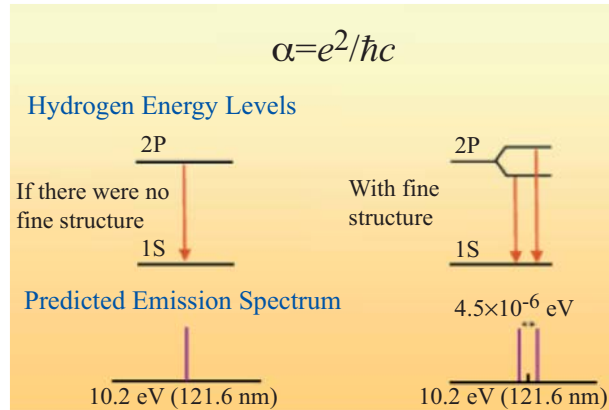

*Fig. 1 - Hydrogen energy levels without and with the hyperfine splitting*

**dance to do on the computer to make this number come out, without putting it in secretly!"**

The idea of reducing everything to numbers is certainly not new, it traces back to Pythagoras and it is now becoming central in the theory of every thing (TOE), whose very first formulation can be found in Plato's Timaeus [3] where he proposed a TEO, employing two right triangles with lengths $(1,1,\sqrt{2}) - (1,\sqrt{3},2)$ respectively. Things have tremendously evolved since that times, but still the fundamental questions are the same[3].

In very recent times Gilson [4] has proposed a simple formula, which yields a remarkable numerical agreement with the experimental value, namely

$$\alpha(\beta) = \frac{\pi}{\beta}\cos(\beta)\,Tanc(\frac{\beta}{29}),$$

$$Tanc(x) = \frac{\tan(x)}{x},$$

$$\beta = \frac{\pi}{137}$$

(1).

---

[3] Ideas like the Higgs field and its associated particle (the Higgs Boson, some times referred as the God particle), the Quintessence, introduced to explain the Dark matter, seem to reflect ancient medieval conceptions. Nowadays the Galilean method has imposed the constraint that pure speculations are nothing if not supported by experiment, notwithstanding, Dirac, one of the fathers of modern Physics, has expressed the following opinion "... it more important to have beauty in one's equations than to have them fit experiments".

The Gilson's formula expresses the fine structure constant in terms of one transcendent number and of two primes[4] (the 10-th and the 33-rd).

The numerical value obtained from eq. (1) is

$$\alpha(\frac{\pi}{137}) \cong \frac{1}{137.035999786699} \qquad (2)$$

Which coincides with the value reported by CODATA 2007 [3] within $3 \cdot 10^{-11}$.

We will not discuss the theoretical assumption behind the derivation of Eq. (1), but, we only remark that it meets the "alchemic" requirements, invoked in the incipit of these introductory remarks. It is also worth noting that its series expansion is very rapidly converging and the first five terms of the expansion,

(Namely $\alpha(\beta) \cong \frac{\pi}{\beta} + \frac{2521}{5046}\pi\beta + \frac{52995607}{254621160}\pi\beta^3 + \frac{761225887831}{8993728613520}\pi\beta^5 + o(\beta^7)$) reproduces the value of $\alpha(\pi/137)^{-1}$ up to the 15-th digit. We have remarked this point for a not secondary reason, we will discuss in the forthcoming parts of the paper.

We have used the adjective alchemic, associated with numerical, not to deny the attitude of putting numbers together without any specific theoretical guidance, but rather to associate this effort with a genuine philosophical and spiritual discipline, which in its original conception, was aimed at penetrating the nature of things, under the guidance of substances possessing unusual properties.

Prime numbers and transcendent numbers are unusual, or, at least, we perceive them as unusual.

Very few people know that Newton, one of the fathers of the modern Physics, was the "last great alchemist" [5] and that milestones of quantum Physics were derived using numerological arguments. The Balmer series was derived indeed by a school master [6] with a non common ability in putting numbers together and the Planck blackbody law was originally obtained (by Planck himself) by means of a fitting procedure [7], in an attempt of reconciling Wien and Jeans laws. Remarkably, the Planck constant emerged from this procedure as a fitting parameter.

---

4   One may ask why these two primes? 137 might be obvious but why 29 ? just to add further elements of speculation we note that they belong to those family of primes of the form $p(n) = n^2 - 7$ and in this list they are consecutive and corresponds to $p(6)$, $p(12)$.

Coming back to the fine structure constant, we remind that Eddington [8] tried to reconcile it with cosmological arguments and proposed the following relationship between $\alpha$ and the number of protons in the universe

$$N_E = 2^{256}\alpha^{-1} \tag{3}$$

This number played an important role in Large Number Hypothesis [9] of central importance in "alternative cosmologies".

We shall reconsider these last points in the concluding section of the paper, here we whish to present a different formula for the fine structure constant, inspired to the alchemic principle implicitly contained in the Feynman's statement. We will add however some physical considerations, helpful to swallow the unjustified steps leading to its formulation.

## 2. FINE STRUCTURE CONSTANT A SIMPLE FORMULA FOR ITS EVALUATION

There is a simple and notable formula yielding $\alpha$ to a very good approximation, it has a Pythagorean flavour and reads [10]

$$\frac{1}{\alpha^2} = \pi^2 + 137^2 = 137^2(1+\beta^2) \tag{4}.$$

The numerical value, we can draw from the above formula, is $\alpha \cong 137.0356$ and we can associate to it a geometrical interpretation, namely $1/\alpha$ is the hypotenuse of a right triangle with catheti $\pi, 137$.

We will not comment any more the previous geometrical remark and only assume that the above relation is essentially correct and make the step further of recovering a complete agreement with the experimental value by considering the following "slight" redefinition

$$\frac{1}{\alpha^2} = 137^2(1+\frac{\beta^2}{1+\beta^{2\phi(\beta)}}),$$

$$\phi(\beta) = \left\{ 2^{r-1}\beta \right\}_{r=1}^{p} \tag{5}.$$

Where $r$ runs on prime numbers and $\{a_m\}_{m=1}^{n}$ denotes the continued fraction expansion

$$\left\{ a_m \right\}_{m=1}^{n} = 1 + \cfrac{a_1}{1 + \cfrac{a_2}{1 + ... 1/a_n}} \qquad (6)$$

By keeping $p=11$ in eq. (5) we find (alc.=alchemic)

$$\alpha_{alc.}^{-1} \cong 137.035999710(27) \qquad (7)$$

which is extremely close to the very accurate evaluation

$$\alpha_{exp}^{-1} \cong 137.03599970(98) \qquad (8)$$

reported in ref. [11].

Before proceeding further let us note that the series expansion of eq. (5) yields

$$\alpha^{-1} \cong \alpha_0^{-1} + \Delta\alpha^{-1}$$
$$\alpha_0^{-1} = 137\, p(\beta) + ...$$
$$\Delta\alpha^{-1} = -137 \cdot \ln(\beta)\,\beta^5 + F(\ln(\beta))\,\beta^6 + G(\ln(\beta))\,\beta^7 + ... \qquad (9)$$
$$p(x) = 1 + \frac{1}{2}x^2 - \frac{685}{2^3 \cdot 137}x^4 + \frac{1781}{2^4 \cdot 137}x^6 + ...$$
$$F(x) = 274\,x - 137\,x^2, G(x) = -\frac{9179}{2}x + 584\,x^2 - \frac{274}{3}x^3$$

In Equation 9 we have indicated two contributions ($\alpha_0^{-1}$, $\Delta\alpha^{-1}$) having different series expansion behaviours

a) the expansion of the part in which $\phi(\beta) = 1$

b) the expansion of the contribution deriving from the dependence of the exponential on $\beta$, they contain terms logarithmically divergent with the expansion parameter.

Even though $\ln(\beta)$, and its successive powers as well, are large numbers, their contribution to the series is well controlled by the product with higher order powers of $\beta$. The terms containing the logarithms appears nested in the series given in eq. (9) as coefficients in the expansion parameter $\beta$.

The series has also two distinct convergence behaviours, the first part converges, in very few terms (the first 4 are sufficient), to $\alpha_0^{-1} \cong 137.035996793137$ *(76)*, the second part has a very slow convergence. With the chosen $p$ value of *11* a very larger number of terms are indeed necessary to obtain the contribution of $\Delta\alpha^{-1} \cong 2.90324041429812 \cdot 10^{-6}$, yielding the $\alpha$ value reported in eq. (7).

The reason we have quoted the behaviour of the series expansions of the Gilson formula and of eq. (5) stems from the fact that we believe that, if a simple formula for $\alpha$ exists, it should be specified by an expansion, which, to some extent, reflects the successive level of approximation implicitly contained in the diagrammatic procedure leading to its theoretical evaluation [11].

The experimental determination of the fine structure constant comes from two distinct steps

a)   the experimental determination of the Dirac electron anomalous magnetic moment, whose dimensionless Landè factor can be written as

$$g = 2(1 + a) \tag{10}$$

b)   the evaluation of the "anomalous" term $a$, through QED corrections at different orders in the fine structure constant, which read

$$a(QED) \cong A_1 + A_2\left(\frac{m_e}{m_\mu}\right) + A_2\left(\frac{m_e}{m_\tau}\right) + A_3\left(\frac{m_e}{m_\mu}, \frac{m_e}{m_\tau}\right) + \dots \tag{11}$$

The various contribution involving the ratio of the electron mass to muon or to tau masses are linked to the following expansion in terms of $\alpha$

$$A_i \cong \sum_{n=1}^{N} A_i^{(2n)}\left(\frac{\alpha}{\pi}\right)^n \tag{12}$$

Just to give an idea the computation of the contribution with *n=4* ,includes the evaluation of 891 Feynman graphs and typical diagrams involved in the process are reported in fig. 2.

In the forthcoming section we will discuss more deeply the meaning of the previous series (11, 12), here we note that we have deviated from the alchemic strategy entering a "too pragmatic" field.

We have defined the function $\phi(\beta)$ as a kind of Padè expansion characterized by powers of 2 raised to $p - 1$, where $p$ is a prime. There is no reasons, other than aesthetical, to make such an assumption. We can however use a different definition and associate the coefficients of the Padè expansion with the so called Euler totient function $\varphi(n)$, defined as the number of integers less or equal to $n$, which are co-primes to $n$, so, for example, $\varphi(6) = 2$.

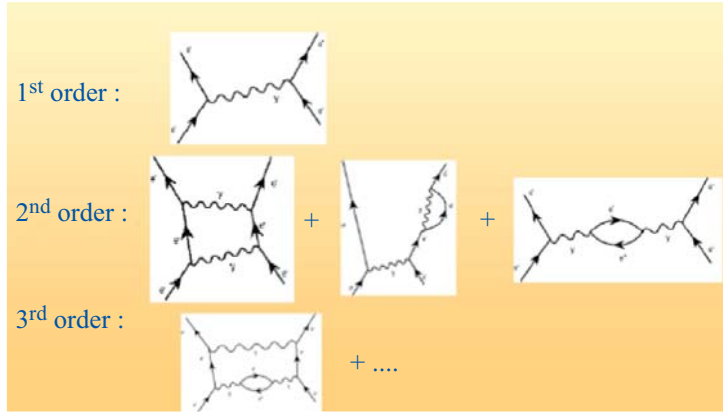

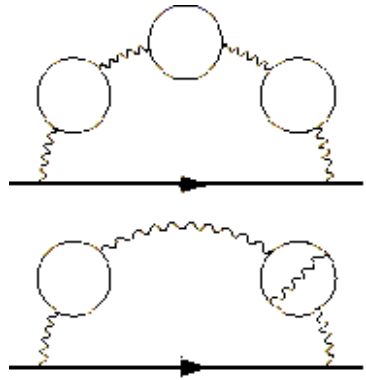

*Fig. 2 - Examples of 1-st, 2$^{nd}$, 3-rd …8-th order Feynman diagrams contributing to the electron self energy. The solid line represents electrons the wavy lines virtual photons, the circles are virtual electron anti-electron pairs*

We note that the following sum defined in terms of the totient function

$$t(n) = \sum_{k=1}^{n} \varphi(k) \tag{13}$$

generates the numbers 0, 1, 2, 4, 6, 10, 12, 18, 22, 28, 32… .

We use, accordingly, the following definition of the $\phi(r)$ function

$$\phi(r) = \left\{ 2^{t(n)} \beta \right\}_{n=0}^{r} \tag{14}$$

Therefore by keeping $r=11$ we obtain $\alpha^{-1} \cong 137.035999710(18)$ which is as good as the value obtained by means of the original prescription.

We can consider unsatisfactory the agreement between alchemic (eq. (7)) and experimental (eq. (8)) values, to this aim we change the definition of the function as it follows

$$\phi(r) = \left\{ 2^{\psi(n)} \beta \right\}_{n=0}^{r} \tag{15}$$

Where

$$\psi(n) = \begin{array}{ll} 0 & n=0 \\ n & such\,that\ \ 6^n + 5 \equiv prime \end{array}$$

$$0,\ 1,\ 2,\ 4,\ 7,\ 10... \tag{16}$$

In this way we get

$$\alpha^{-1}_{alc.} \cong 137.03599970(90) \tag{17}$$

The correspondence can be made even more precise by keeping the successive terms in the series.

If we are not satisfied with this last definition of the function $\psi(n)$, we can propose as further alternative

$$\psi(n) = \begin{array}{ll} 0 & n=0 \\ n & divisor\,of\ \displaystyle\sum_{m=1}^{n} m^n \end{array}$$

$$0,1,2,4,7,10,14,28... \tag{18},$$

which yields results closely similar to those shown in eq. (17).

The above examples are purely alchemic combinations, which do not meet any scientific criterion and the only indication they give is that the values of the fine structure constant are reproducible by an equation of the type (4) in which the exponent formula is expressible by a finite continued fraction.

In this section we have shown that the fairly simple function (5) yields a god approximation of the experimental result reported in ref. [11] and obtained by comparing the experimental value of the electron anomalous magnetic moment with the QED corrections, calculated with an awkwardly complex procedure, involving the computations of a huge amount of Feynman diagrams and with an amazingly amount of computer time.

We have also stressed that the associated series expansion reflects, to a certain extent, a kind of diagrammatic expansion involving higher orders loop diagrams, this is the reason why we believe that formulae which define α to great precisions but with a naïve series expansion in terms of β, can be ruled out.

The Gilson formula does not posses these features. It has indeed been worked out to get read of any diagrammatic expansion and its uncertainties, associated with the renormalization methods. The experimental determination should therefore a non QED diagrammatic procedure, using therefore a direct method, namely a quantity depending only on $\alpha$. The Quantum Hall effect [12], as also suggested by Gilson, seems to be an ideally suited candidate.

## 3.    ALFA CONSTANT AND THE HALL EFFECT

The Hall effect [13] is well known since more then one century and is sketched in Fig. 3. When a magnetic field is applied perpendicularly to the direction of flowing of a current in a metal, a voltage is developed in the third direction.

The potential is just due to the deflection of positively and negatively charged carriers towards the edge of the metal sample. The Hall potential $V_H$ can be easily calculated at equilibrium, namely the magnetic force is balanced by the electric force, and reads

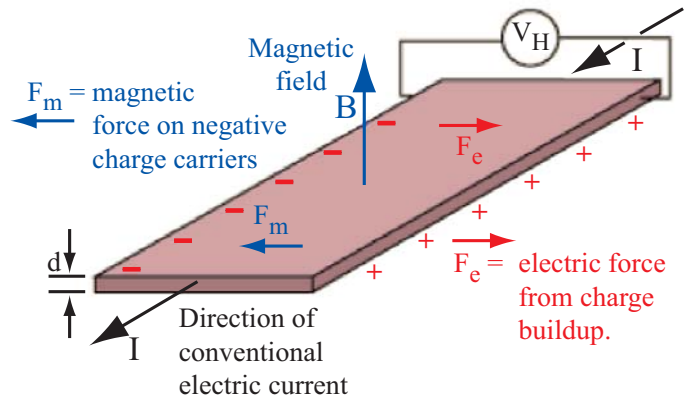

*Fig. 3 - Sketch of the Hall effect*

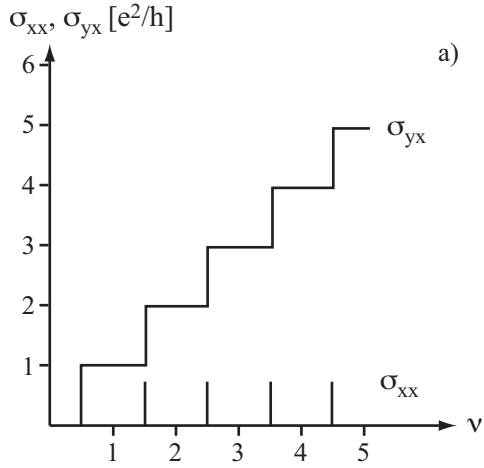

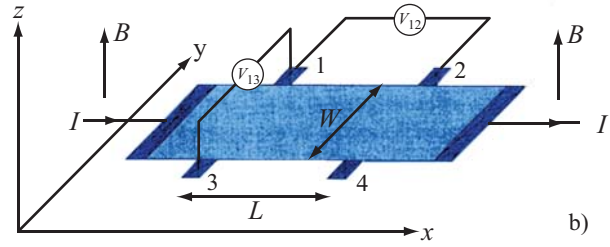

*Fig. 4a - Quantum Hall resistivities vs. the magnetic field intensity*

*Fig. 4b - Definition of the quantum Hall effect resistivities $\sigma_{x,y} \propto V_{1,3}/I = R_H$, $\sigma_{x,x} \propto V_{1,2}$*

$$V_H = I \frac{B}{n\,e\,d} \tag{19}.$$

The associated Hall resistance $R_H = B/ned$ increases monotonically, with he applied magnetic field.

The quantum Hall effect occurs at low temperature and is a manifestation of the so called Landau quantization. In a two dimensional metal (and in a semiconductor as well) the Hall resistance does not continuously increase with increased magnetic field, but exhibits the step behaviour shown in Figs. 4. It is quantized and expressed in terms of a fundamental resistance

$$r_H^* = \left(\frac{e^2}{h}\right)^{-1} \cong 25181.2\,\Omega \tag{20}$$

known as the von Klitzing constant and used to calibrate very accurately the resistance [14].

It is evident that the Klitzing is directly associated with the fine structure constant.

The measurements done with the quantum Hall effect are amazingly precise and they may allow a diagrammatic independent determination of the fine structure constant. The obtained results corroborates those obtained with the procedure described in ref. [11], but cannot be considered more precise since they are affected by other uncertainties connected with those associated with the light velocity and the laboratory calibration of the resistances.

# 4. PADÈ AND NON PADÈ APPROXIMANTS AND QED SERIES

Before concluding this paper, it is worth spending a few words on series computation technicalities, which are not totally extraneous to the alchemic principle, which to some extent, has inspired the present analysis.

The method of Feynman diagrams is an efficient tool, since it allows the evaluation of processes in QED with extremely high accuracy, the involved calculations are however extremely cumbersome and some times there are not simple procedures to understand the value of a contribution or even its sign, without performing an enormous amount of computations.

Padè approximants (PA) [15] have been recognized since long times as powerful tools, both in applied Mathematics and Physics, to deal with a broad range of problems, involving perturbative expansions, since they are known to accelerate the convergence of a series.

The perturbative expansions of QED, and of QCD as well, are therefore a natural field of application of the PA [16,17].

The usually adopted procedure is that of considering a given QED series, evaluate the associate PA and determine the value of the total series or the next term in the expansion. The validity of the procedure should be used with extreme caution, because in most cases works, but it is not clear why.

We remind that a PA of order $m + n$ is denoted by $[m \mid n]$ and is used to indicate the following ratio between two polynomials of degree $m$ and $n$ respectively

$$[m \mid n] = \frac{P_m(x)}{Q_n(x)} \tag{21}.$$

To understand how PA works we consider the function

$$f(x) = \frac{\ln(1 + x)}{x} \tag{22}$$

whose third order series expansion around the origin is given by

$$f(x) \cong 1 - \frac{x}{2} + \frac{x^2}{3} \tag{23}$$

according to eq. (21) a second order PA approximant is

$$[1|1] = \frac{a_0 + a_1 x}{1 + b_1 x} \qquad (24).$$

We can evaluate the coefficients $a, b$ by matching the second order expansions of (24) and (22). We obtain therefore

$$a_0 = 1, a_1 = \frac{2}{3}, b_1 = \frac{1}{6}$$

Equation 24 can be now confronted with the original function (as shown in Fig. 5) where it has been shown that, albeit we used two terms of the Taylor expansion only we have obtained a much better approximation. The figure does not report the Taylor expansion approximation, which fails for $x \le 0.5$.

In ref. [17] it has been shown by the present Author that Non Rational Padè Approximants (NRPA) may, in many cases, provide better approximations[5].

A first order NRPA reads

$$E[1|x] = (a_0 + a_1 x)^{\beta} \qquad (25)$$

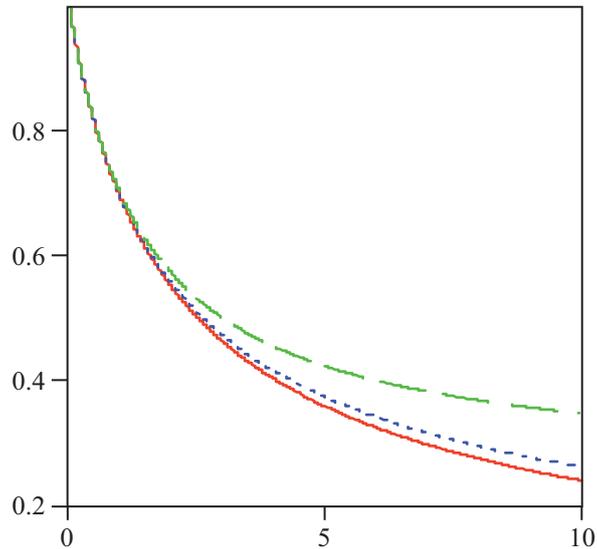

*Fig. 5 - Comparison between eq. (22) (continous), PA (dash), NRPA (dot)*

---

[5] The Author recognizes that the theory of non rational Padè approximants was originally developed in collaboration with A. Segreto as a tool of analysis for the theory of Free Electron Laser.

the same prescription as before yields for the coefficients of the NRPA (25) the following values

$$a_0 = 1, a_1 = \frac{5}{6}, \beta = -0.6 \qquad (26)$$

The comparison between the various expansions is reported in Fig. (5), and it is evident that the NRPA yields a much better agreement even for $x \cong 10$.

Second order NRPA have been reported in ref. [17] and they provide much better approximations compared to higher orders rational PA.

Such a method allows to cast the QED expansion reported in eq. 12 in the form

$$A_1 \cong \sum_{n=1}^{4} A_1^{(2n)} x^n \cong \frac{x}{2} \Big[ 1 + a_1 x + a_2 x^2 \Big]^{\beta},$$
$$A_1^{(2)} = \frac{1}{2}, x = \frac{\alpha}{\pi} \qquad (27)$$

and the coefficients $a_{1,2}, \beta$ are simple functions of the coefficients $A_1^{(2n)}$ reported in ref. [11] and whose evaluation requires [18]

| | | | |
|---|---|---|---|
| *n=2* | *7* | *Feynman diagrams* | $A_1^{(4)} = -0.32847896$ |
| *n=3* | *72* | " | $A_1^{(6)} = 1.1811241456$ |
| *n=4* | *891* | " | $A_1^{(8)} = -1.7283$ |

The contribution with *n=5*, not yet calculated, would require the computation of 12672 Feynman diagrams.

We can attempt a prediction of this further contribution, by writing the second order NRPA using the data up to *n=4*, thus finding

$$A_1 = \frac{1}{2} x \Big[ 1 + 0.980439953429742 \cdot x - 2.72271965268675 \cdot x^2 \Big]^{-0.670064411085913} \qquad (28)$$

The expansion of the term in the square bracket up to the 4-th order yields, for the contribution *n=4*, the value $A_1^{(10)} \cong 2.24008$.

The example we have reported in this section has just been aimed at providing the hint that even at this level, fairly elaborated of approximation, elements of a bizarre way of thinking can still be envisaged.

## 5. AN OBSESSION BUT WHY? THE DILEMMA 137 OR SOMETHING ELSE

The last century has been the century of great revolutions and of great mystifications.

It started with the Planck hypothesis of light quanta and in twenty years the Physics itself was shocked and changed in its foundations.

Wolfgang Pauli has been one of the most genuine interpreter of the new era. He was born just in 1900 and dominated the Physics for more then thirty years. He was the member of a distinguished intellectual family of German-Hebrew culture, converted to Catholicism and his good-father was Ernest Mach.

It is certainly a notable feature that Pauli and Carl Gustav Jung had a cooperation and a correspondence on different topics ranging from Physics to Psychology.

Even though it is the personal opinion of the author that certain aspects of psychology and the introduction of psychoanalysis belong to the great mystifications of the century, it is worth to stress that part of this cooperation was motivated by the analysis of the of Pauli's dreams, whose profound motivation can be traced back to Alchemy and Cabbala. Jung was the Author of the book *Psychology and Alchemy* [18] and the fine structure constant was one of the subjects of Pauli's obsession.

The topics covered by the Pauli-Jung cooperation are so vast and profound that a paragraph in a semi-technical article cannot even grasp the surface, a recent and authoritative article on the subject will be, for the interested reader, a concrete source of further inspiration [20][6].

However we want to remark that the Pauli's interest for $\alpha$ lasted since the very beginning of his scientific career, being a Sommerfeld's pupil, he was initially very much impressed by the points of view of his mentor and of Rydberg as well.

---

[6]  The paper is interesting for various reasons, among the other things it contains the following interesting approximation of the fine structure constant $\alpha^{-1} \cong 4\pi^3 + \pi^2 + \pi$

In his book Atombau and Spektrallienen [21], Sommerfeld defined the so called Rydberg top square equation a Cabbalistic formula and perhaps these early suggestions, along with his Catholic-Hebrew heritage contributed to his future conceptions.

The following Pauli's comments extracted from his (philosophical or non technical) essays may be illuminating regarding the previous points

> *In a big paper "Untersuchungen über das system der Grundsoffe" of 1913 he goes one step further. After the quotation of the earlier formulas 2=2×1$^2$, 8=2×2$^2$ and 18=2×3$^2$ he goes...: "the continuation would be 2×4$^2$=32, 2×5$^2$=50 etc." This is the famous formula 2×p$^2$ (p integer) which Sommerfeld called "cabbalistic" in this book "Atombau und Spectrallinien" the group G4 "(p=4, rare earths) that it consist of 32, not of 36 elements".*

> *The series of whole numbers 2, 8, 18, 32... giving the lengths of the periods in the natural system of chemical elements, was zealously discussed in Münich, including the remark of the if "n" takes on all integer values. Sommerfel tried especially to connect the number 8 and the number corners of a cube".*

> *"A new phase of my scientific life began when I met Niels Bohr personally for the first time. This was in 1922, when he gave a series of guest lectures at Göttingen, in which reported on his theoretical investigations on the periodic system of elements. I shall recall only brief that the essential progress spherical symmetric atomic model....".*

The following remark is extremely important for the purposes of the present discussion

> *From the view of logic my report on "Exclusion principle and quantum mechanics" has no conclusion. I believe that it will only be possible to write the conclusion if a theory will be established which will determine the value of the fine structure constant and will thus explain the atomistic of electric fields actually occurring in nature".*

The above selected comments have been taken from reports relevant to different periods of Pauli's scientific life. The last, from his Nobel prize lecture, seems to report a feeling of frustration for having not succeeded, in thirty years of work, in reconciling everything in a self contained and unitary vision of the nature.

The leit motiv of all the above considerations is, however, a kind of thread (alchemical, numerological or whatever) which has certainly played a not secondary role in the scientific path, bringing Pauli to the formulation of the exclusion principle.

A further element of fascination was the link, implicitly contained in the definition of the fine structure constant, which is that of two quanta, namely the Planck constant and the charge of the electrons.

In an essay dedicated to the Einstein contributions to Quantum mechanics, Pauli expressed the following opinion

*Inside physics in the proper sense we are well aware that the present edifice of quantum mechanics is still far from its final form, but, on the contrary, leaves problems open which Einstein considered already long ago. In this previously cited paper of 1909.... He stresses the importance of Jeans' remark that the elementary electric charge e, with the help of the velocity of the light c, determines the constant $e^2/c$ which is of the same dimensions as the quantum of actions h (thus aiming at the now well known fine structure constant $2\pi e^2/hc$). He emphasized "that the elementary quantum of electric e is a stranger in Maxwell-Lorentz' electrodynamics" and expressed the hope that "the same modifications of the theory which contains the elementary quantum e as a consequence, will also have as a consequence the quantum structure of radiation." The reverse of this statement certainly turned out to be not true, since the new quantum theory of radiation and matter does not have the value of the elementary electric charge as a consequence, so that the latter is still a stranger in quantum mechanics too.*

*The theoretical determination of the fine structure constant is certainly the most important of the unsolved problems of modern physics. To reach it, we shall, presumably, have to pay with further revolutionary changes of the fundamental concepts of physics with a still farther digression from the concepts of the classical theories".*

Analogous concepts were stressed in one of his last papers [22]

*"One of the most assured empirical results of physics is the atomistic structure of electric charge. Charge values are integral multiples of a fundamental unit, the electric elementary quantum, from which, along with the quantum of action and velocity of light, one can from a dimensionless number, 137.04. To reach this result one requires a considerable part of he classical theory of electricity. In the 17$^{th}$ century, for instance, when it was not known how to measure electric charges and how they are defined quantitatively, this empirical result could never have been obtained and formulated. But we are unable to understand or explain the above number".*

The quantization of charge finds a "natural" explanation in the theory of magnetic monopoles developed by Dirac [23]. We remind therefore that if a monopole exists it would be subject to a force, analogous to the force due to the electric field on an ordinary charge, namely

$$F_M = Q_M B \tag{29}$$

where $Q_M$ is a magnetic charge, by multiplying "electric and magnetic forces we find

$$F_E F_M = Q_m Q_E E B \tag{30}.$$

The use of simple dimensional arguments shows that the product of the electric and magnetic charges has the dimensions of an action. It is therefore not strange that the quantization of the charge passes through the following quantization rule (CGS units)

$$\frac{Q_m Q_e}{2\pi} = n\hbar \tag{31}.$$

By assuming that $Q_m = m\,g, Q_e = q\,e$, where $m,\ q$ are integers and $g, e$ are the elementary magnetic and electric charge respectively, we find ($m=q=1$)

$$\alpha\,g \propto n\,e\,c \tag{32}.$$

The point of view raised by Pauli had therefore a sound motivation. On the other hand an alternative way of writing eq. (32) is

$$\frac{g}{e} \propto n\,\frac{h}{e^2} = n\,R_H \tag{33}$$

Which is the quantization emerging from the quantum Hall effect.

Arguments related to the Dirac monopoles may allow the derivation of independent formulae for the definition of the fine structure constant. One example reported in ref. [23] is the Wyler "marvellous" formula (originally proposed in [25])

$$\alpha = \frac{9}{16\pi^3} \sqrt[4]{\frac{\pi}{5!}} \tag{34}.$$

All the discussion of this paper and of the associated references have turned around the point that $\alpha^{-1}$ is extremely close to 137 with all its cabalistic meanings[7], but we know that this value is a low energy approximation, the values of the coupling constants (strong, electromagnetic and weak) are depending on energy, as shown in Fig. 6 and they tend at the same value at very high energies.

The fine structure constants at the energy scale of the *W boson (81 GeV)* is 1/128, it seems therefore that the fundamental character of *137* looses any meaning.

This is not a conclusion, but just an open question.

The author of this paper is unfortunately not too much bent towards mysticism or conjectures, going beyond the laboratory size scale, he may have therefore treated this topic without the necessary sensitivity and cultural preparation. These deficiencies may have lead him to underestimate important aspects and implications of the underlying subtleties. Notwithstanding he remains convinced that the value of the $\alpha$ coupling constant, the number we measure or, better, we observe, is just what it is and cannot be different.

---

[7]   In ancient Hebraic language letters where used for numbers, and Cabbala is the word corresponding to *137.*

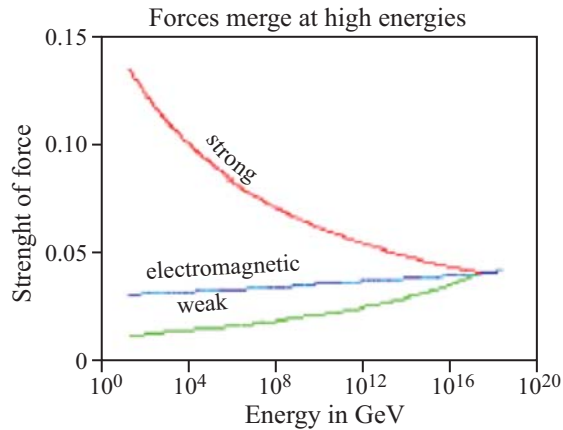

*Fig. 6 - Coupling constants of the fundamental forces vs energy.*

The prize for a different value could be that the "biochemical Observer" pondering on the origin of universe could not exist.

Life[8], at least in the form we know it, requires a universe sufficiently old to be sufficiently cold. Complex aggregations, at the basis of the life itself, would not be made possible in a more energetic environment[9].

The impression is therefore that asking why the fine structure constant has precisely that value is just a restatement of the question why are we here?

Frankly speaking, not too much progress can be expected in this direction in the next, medium and far future.

Furthermore, on the top of that, the Gödel's theorem [25] should not be forgotten. We can summarize it using what is referred as the Gödel's sentence"

**There are statements which are true, but cannot be proven**

---

[8] Non biochemical observers, namely different forms of intelligence supported by forms of life not based on biochemical aggregates and therefore not relying upon the electromagnetic interaction, could be hypothesized too, but this is too speculative to be taken much seriously at the moment.

[9] This last statement could be used as a contra-argument ascribing to the low Energy limit an anthropocentric role and therefore to 137 an alchemic role.

It might be discouraging, but we strongly feel that it applies to the case discussed in this article.

This is certainly at the opposite side of the Hilbert program based on the assumption

**We must know-we will know[10]**

Some time there is a feeling of frustration when what we believe the essence (real or supposed) of things is far away from our understanding. Much more then words, images can communicate such a feeling and I am sure that the Magritte painting reported in Fig. 7 is an appropriate conclusion for this article.

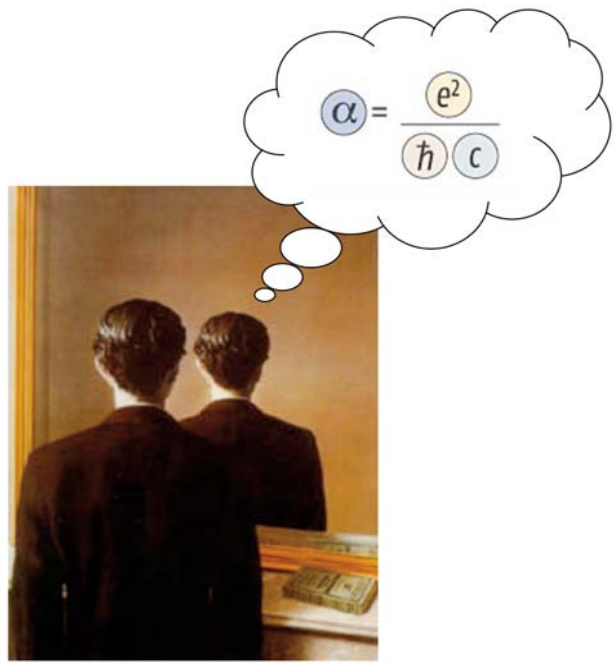

*Fig. 7 - R. Magritte. La reproduction interdite with a personal modification of the Author*

**ACKNOWLEDGMENTS**

It is a pleasure to thank Drs. M. Quattromini and M. Del Franco for their kind assistance and suggestions.

---

[10] In German it sounds even more categorical "Wir Mussen wissen-Wir werden wissen" and it is the epitaph on the Hilbert's tomb in Konigsberg, for a strange joke of the destiny this sentence was pronounced just one day before that in which Gödel presented his thesis, containing his theorem.